\documentclass{elsart}
\usepackage{amsfonts}
\usepackage{amsmath}
\usepackage{amssymb}
\usepackage{bm}
\usepackage{graphicx}

\input{tcilatex}

\begin{document}

\begin{frontmatter}

\title{New eigen-mode of spin oscillations in the triplet superfluid condensate in neutron stars}

\author{L. B. Leinson}
\address{Institute of Terrestrial Magnetism, Ionosphere and
 Radio Wave Propagation RAS, 142190 Troitsk, Moscow Region, Russia}

\begin{abstract}
The eigen mode of spin oscillations with $\omega\simeq \sqrt{58/35}\Delta$  is predicted to exist besides already known spin waves with $\omega \simeq\Delta /\sqrt{5}$ in the triplet superfluid neutron condensate in the inner core of neutron stars. 
The new mode is kinematically able to decay into neutrino pairs through neutral weak currents. 
The problem is considered in BCS approximation for the case of $^{3}P_{2}-$\noindent$^{3}F_{2}$ pairing with a projection of the total angular momentum $m_{j}=0$ which is conventionally considered as preferable one at supernuclear densities.
\end{abstract}


\end{frontmatter}

A superfluidity of the inner core of neutron stars plays a crucial role in
theirs cooling scenario. The energy gap $\Delta $ arising in the
quasiparticle spectrum below the critical condensation temperature $T_{c}$
suppresses the most of neutrino emission mechanisms \cite{YL}. According to
the minimal cooling paradigm \cite{Page04,Page09,SY,PPLS}, under these
conditions, the most efficient energy losses from the star volume can take
place at a recombination of thermal excitations in the form of broken Cooper
pairs. The neutrino emission at the pair-recombination processes occurs
through neutral weak currents in the axial channel of weak interactions%
\footnote{%
The vector channel of the neutrino radiation through neutral weak currents
is strongly suppressed in the non-relativistic case \cite{LP06,L09a}.} and
can be very efficient, in the triplet superfluid neutron liquid, a few below
the critical temperature \cite{YKL}. However, the corresponding neutrino
emissivity falls rapidly with lowering of the temperature because the number
of broken pairs, having the excitation energy larger than $2\Delta $,
decreases exponentially. In this case the \emph{collective} excitations of
the condensate can dominate in the neutrino production.

Since we assume that the condensate consist of neutron pairs in the triplet
state it is natural to expect the collective modes associated with spin
fluctuations of the condensate\footnote{%
Previously spin modes have been thoroughly studied in the $p$-wave
superfluid liquid $^{3}He$ with a central interaction between quasiparticles 
\cite{Maki,C1,C2,W,Wolfle} . These results cannot be applied directly to the
triplet-spin neutron superfluid condensate, where the pairing is caused
mostly by the spin-orbit interaction between quasiparticles (see details in
Ref. \cite{L10a})}. Such collective excitations with the energy lower than $%
2\Delta $ might undergo the weak decay into neutrino pairs. Recently spin
waves with the excitation energy $\omega =\Delta /\sqrt{5}$~was predicted to
exist in the superfluid spin-triplet condensate of neutrons \cite%
{L09a,L10a,L10b}. Because of a rather small excitation energy, the weak
decay of such waves leads to a substantial neutrino emission at the lowest
temperatures $T\ll T_{c}$, when all other mechanisms of the neutrino energy
losses are killed by the superfluidity.

In Refs. \cite{L09a,L10a,L10b}, the eigen-mode of spin oscillations in the $%
^{3}P_{2}$ superfluid neutron liquid was studied in a simple model
restricted to excitations of the condensate with $l=1$. In this paper we
demonstrate that extending of the decomposition up to $l=1,3$ leads to a
very small frequency shift of the known mode, $\omega =\Delta /\sqrt{5}$,
but opens the new additional mode of spin oscillations with the finite
energy gap $\omega \left( \mathbf{q}=0\right) <2\Delta $. The problem is
considered for the case of $^{3}P_{2}-$\noindent $^{3}F_{2}$ pairing with a
projection of the total angular momentum $m_{j}=0$ which is conventionally
considered as preferable one at supernuclear densities.

We will examine the spin modes within the BCS approximation\footnote{%
Throughout this paper, we use the system of units $\hbar =c=1$ and the
Boltzmann constant $k_{B}=1$.}. Let us remind briefly the theory of spin
density excitations in the condensate. The order parameter, $\hat{D}\equiv
D_{\alpha \beta }$, arising due to triplet pairing of quasiparticles,
represents a $2\times 2$ symmetric matrix in spin space, $\left( \alpha
,\beta =\uparrow ,\downarrow \right) $. The spin-orbit interaction among
quasiparticles is known to dominate in the nucleon matter of a high density.
Therefore it is conventional to represent the triplet order parameter of the
system as a superposition of standard spin-angle functions of the total
angular momentum $\left( j,m_{j}\right) $, 
\begin{equation}
\Phi _{\alpha \beta }^{\left( j,l,m_{j}\right) }\left( \mathbf{n}\right)
\equiv \sum_{m_{s}+m_{l}=m_{j}}\left( \frac{1}{2}\frac{1}{2}\alpha \beta
|sm_{s}\right) \left( slm_{s}m_{l}|jm_{j}\right) Y_{l,m_{l}}\left( \mathbf{n}%
\right) .  \label{sa}
\end{equation}%
Assuming that the pair condensation occurs into the state with a total
angular momentum $j=2$ we use the vector notation which involves a set of
mutually orthogonal complex vectors $\mathbf{b}_{l,m_{j}}\left( \mathbf{n}%
\right) $ defined as%
\begin{equation}
\mathbf{b}_{l,m_{j}}\left( \mathbf{n}\right) =-\frac{1}{2}\mathrm{Tr}\left( 
\hat{g}\bm{\hat{\sigma}}\hat{\Phi}^{2,l,m_{j}}\right) ~,~\mathbf{b}%
_{l,-m_{j}}=\left( -\right) ^{m_{j}}\mathbf{b}_{l,m_{j}}^{\ast },
\label{blm}
\end{equation}%
where $\bm{\hat{\sigma}}=\left( \hat{\sigma}_{1},\hat{\sigma}_{2},\hat{\sigma%
}_{3}\right) $ are Pauli spin matrices, $\hat{g}=i\hat{\sigma}_{2}$, and the
angular dependence of the order parameter is represented by the unit vector $%
\mathbf{n=p}/p$ which defines the polar angles $\left( \theta ,\varphi
\right) $ on the Fermi surface. The vectors $\mathbf{b}_{l,m_{j}}$ are
mutually orthogonal and are normalized by the condition%
\begin{equation}
\left\langle \mathbf{b}_{l^{\prime },m_{j}^{\prime }}^{\ast }\mathbf{b}%
_{l,m_{j}}\right\rangle =\delta _{ll^{\prime }}\delta _{m_{j}m_{j}^{\prime
}}.  \label{lmnorm}
\end{equation}%
Hereafter the angle brackets denote angle averages, $\left\langle
...\right\rangle \equiv \left( 4\pi \right) ^{-1}\int d\mathbf{n}...$.

The block of interaction diagrams irreducible in the channel of two
quasiparticles, $\Gamma _{\alpha \beta ,\gamma \delta }$, is usually
generated by expansion over spin-angle functions. The spin-orbit interaction
among quasiparticles is known to dominate at high densities. This implies
that the spin $\mathbf{s}$ and orbital momentum $\mathbf{l}$ of the pair
cease to be conserved separately, and the complete list of channels includes
the pair states with $j=0,1,2$, and $\left\vert m_{j}\right\vert \leq j$.
These nine complex states exhaust the number of independent components in
the matrix order parameter arising at the $P$-wave pairing caused by the
strong spin-orbit forces. The pairing in the $j=2$ channel dominates, and
due to relatively small tensor components of the neutron-neutron interaction
the condensation of pairs occurs in the $^{3}P_{2}+$\noindent $^{3}F_{2}$
state. In this pairing model, contributions from $^{3}P_{2}\rightarrow $%
\noindent $^{3}P_{0}$ or $^{3}P_{2}\rightarrow $\noindent $^{3}P_{1}$
transitions are deemed to be unimportant. Such assumption is somewhat
vulnerable especially when considering excited state of the condensate.
Unfortunately the detailed information on the in-medium effective
interaction between neutrons in the channels $j=0,1$ is currently
unavailable and requires a special investigation. Hence we take the
approximation to neglect the $j=0,1$ coupling throughout this paper. From
now on we omit the suffix j everywhere by assuming that the interaction
occurs in the state with $j=2$. Thus we assume $l=j\pm 1$, and 
\begin{equation}
\varrho \Gamma _{\alpha \beta ,\gamma \delta }\left( \mathbf{p,p}^{\prime
}\right) =\sum_{l^{\prime }lm_{j}}\left( -1\right) ^{\frac{l-l^{\prime }}{2}%
}V_{ll^{\prime }}\left( p,p^{\prime }\right) \left( \mathbf{b}_{lm_{j}}(%
\mathbf{n})\bm{\hat{\sigma}}\hat{g}\right) _{\alpha \beta }\left( \hat{g}%
\bm{\hat{\sigma}}\mathbf{b}_{l^{\prime }m_{j}}^{\ast }(\mathbf{n}^{\prime
})\right) _{\gamma \delta },  \label{ppint}
\end{equation}%
where $V_{ll^{\prime }}\left( p,p^{\prime }\right) $ are the interaction
amplitudes, and $l,l^{\prime }=1,3$, in the case of tensor forces; $\varrho
=p_{F}M^{\ast }/\pi ^{2}$ is the density of states near the Fermi surface in
the normal state. The effective mass of a neutron quasiparticle is defined
as $M^{\ast }=p_{F}/\upsilon _{F}$, where $\upsilon _{F}\ll 1$ is the Fermi
velocity of the non-relativistic neutrons.

The order parameter is of the following general form 
\begin{equation}
\hat{D}\left( \mathbf{n}\right) =\sum_{lm_{j}}\Delta _{l,m_{j}}\left( %
\bm{\hat{\sigma}}\mathbf{b}_{l,m_{j}}\right) \hat{g}.  \label{Dnlm}
\end{equation}

The ground state occurring in neutron matter has a relatively simple
structure (unitary triplet) \cite{Tamagaki,Takatsuka}, where 
\begin{equation}
\sum_{lm_{j}}\Delta _{l,m_{j}}\mathbf{b}_{l,m_{j}}\left( \mathbf{n}\right)
=\Delta ~\mathbf{\bar{b}}\left( \mathbf{n}\right) .  \label{bbar}
\end{equation}%
On the Fermi surface, $\Delta $ is a complex constant, and $\mathbf{\bar{b}}%
\left( \mathbf{n}\right) $ is a real vector which we normalize by the
condition 
\begin{equation}
\left\langle \bar{b}^{2}\left( \mathbf{n}\right) \right\rangle =1.
\label{Norm}
\end{equation}%
The following orthogonality relations are also valid:%
\begin{equation}
\int \frac{d\varphi }{2\pi }\mathbf{b}_{l,m_{j}}^{\ast }\mathbf{b}%
_{l^{\prime },m_{j}^{\prime }}=\delta _{m_{j}m_{j}^{\prime }}\mathbf{b}%
_{l,m_{j}}^{\ast }\mathbf{b}_{l^{\prime },m_{j}},  \label{bbdfi}
\end{equation}%
\begin{equation}
\int \frac{d\varphi }{2\pi }\left( \mathbf{\bar{b}b}_{l,m_{j}}^{\ast
}\right) \left( \mathbf{\bar{b}b}_{l^{\prime },m_{j}^{\prime }}\right)
=\delta _{m_{j}m_{j}^{\prime }}\left( \mathbf{\bar{b}b}_{l,m_{j}}^{\ast
}\right) \left( \mathbf{\bar{b}b}_{l^{\prime },m_{j}}\right) .
\label{bbbbdfi}
\end{equation}%
Thus the triplet order parameter can be written as 
\begin{equation}
\hat{D}\left( \mathbf{n}\right) =\Delta \mathbf{\bar{b}}\bm{\hat{\sigma}}%
\hat{g}.  \label{Dn}
\end{equation}

Making use of the adopted graphical notation for the ordinary and anomalous
propagators, $\hat{G}=\parbox{1cm}{\includegraphics[width=1cm]{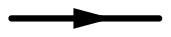}}$, $%
\hat{G}^{-}(p)=\parbox{1cm}{\includegraphics[width=1cm,angle=180]{Gn.eps}}$, 
$\hat{F}^{(1)}=\parbox{1cm}{\includegraphics[width=1cm]{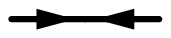}}$\thinspace
, and $\hat{F}^{(2)}=\parbox{1cm}{\includegraphics[width=1cm]{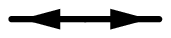}}$%
\thinspace , it is convenient to employ the Matsubara calculation technique
for the system in thermal equilibrium. Then the analytic form of the
propagators is as follows \cite{AGD,Migdal}%
\begin{align}
\hat{G}\left( p_{\eta },\mathbf{p}\right) & =G\left( p_{\eta },\mathbf{p}%
\right) \delta _{\alpha \beta }~,\ \ \ \ \ \ \ \hat{G}^{-}\left( p_{\eta },%
\mathbf{p}\right) =G^{-}\left( p_{\eta },\mathbf{p}\right) \delta _{\alpha
\beta },  \notag \\
\hat{F}^{\left( 1\right) }\left( p_{\eta },\mathbf{p}\right) & =F\left(
p_{\eta },\mathbf{p}\right) \mathbf{\bar{b}}\bm{\hat{\sigma}}\hat{g}~,\ \ \ 
\hat{F}^{\left( 2\right) }\left( p_{\eta },\mathbf{p}\right) =F\left(
p_{\eta },\mathbf{p}\right) \hat{g}\bm{\hat{\sigma}}\mathbf{\bar{b}},
\label{GF}
\end{align}%
where the scalar Green's functions are of the form $G^{-}\left( p_{\eta },%
\mathbf{p}\right) =G\left( -p_{\eta },-\mathbf{p}\right) $ and%
\begin{equation}
G\left( p_{\eta },\mathbf{p}\right) =\frac{-ip_{\eta }-\varepsilon _{\mathbf{%
p}}}{p_{\eta }^{2}+E_{\mathbf{p}}^{2}}~,\ F\left( p_{\eta },\mathbf{p}%
\right) =\frac{-\Delta }{p_{\eta }^{2}+E_{\mathbf{p}}^{2}}.  \label{GFc}
\end{equation}%
Here $p_{\eta }\equiv i\pi \left( 2\eta +1\right) T$ with $\eta =0,\pm 1,\pm
2...$ is the Matsubara's fermion frequency, and $\varepsilon _{\mathbf{p}%
}=p^{2}/\left( 2M^{\ast }\right) -p_{F}^{2}/\left( 2M^{\ast }\right) $. $%
\allowbreak \allowbreak $The quasiparticle energy is given by $E_{\mathbf{p}%
}^{2}=\varepsilon _{\mathbf{p}}^{2}+\Delta ^{2}\bar{b}^{2}\left( \mathbf{n}%
\right) $, where the (temperature-dependent) energy gap, $\Delta \bar{b}%
\left( \mathbf{n}\right) $, is anisotropic. In the absence of external
fields, the gap amplitude $\Delta $ is real.

Finally we introduce the following notation used below. We designate as $%
\mathcal{I}_{XX^{\prime }}\left( \omega ,\mathbf{n,q};T\right) $ the
analytical continuations onto the upper-half plane of complex variable $%
\omega $ of the following Matsubara sums:%
\begin{equation}
\mathcal{I}_{XX^{\prime }}\left( \omega _{\kappa },\mathbf{n,q};T\right)
\equiv T\sum_{\eta }\frac{1}{2}\int_{-\infty }^{\infty }d\varepsilon _{%
\mathbf{p}}X\left( p_{\eta }+\omega _{\kappa },\mathbf{p+}\frac{\mathbf{q}}{2%
}\right) X^{\prime }\left( p_{\eta },\mathbf{p-}\frac{\mathbf{q}}{2}\right) ,
\label{Ixx}
\end{equation}%
where $X,X^{\prime }\in G,F,G^{-}$, and $\omega _{\kappa }=2i\pi T\kappa $
with $\kappa =0,\pm 1,\pm 2...$.These are functions of $\omega $, $\mathbf{q}
$ and the direction of the quasiparticle momentum $\mathbf{p}=p\mathbf{n}$.

We will focus on the processes with $\omega ^{2}<2\Delta ^{2}\bar{b}^{2}$
and with a time-like momentum transfer, $q^{2}<\omega ^{2}$. In this case
the key role in the response theory belongs to the loop integral $\mathcal{I}%
_{FF}$. A straightforward calculation yields $\mathcal{I}_{FF}\left( \mathbf{%
n},\omega ,\mathbf{qn};T\right) =\mathcal{I}_{0}\left( \mathbf{n,}\omega
;T\right) +O\left( q^{2}\upsilon _{F}^{2}/\omega ^{2}\right) $, where 
\begin{equation}
\mathcal{I}_{0}\left( \mathbf{n,}\omega ;T\right) =\int_{0}^{\infty }\frac{%
d\varepsilon }{E}\frac{\Delta ^{2}}{4E^{2}-\left( \omega +i0\right) ^{2}}%
\tanh \frac{E}{2T}.  \label{FFq0}
\end{equation}%
Insofar as $q^{2}\upsilon _{F}^{2}/\omega ^{2}\ll 1$ and $q^{2}\upsilon
_{F}^{2}/\Delta ^{2}\ll 1$ we will neglect everywhere small corrections
caused by a finite value of space momentum $\mathbf{q}$.

The gap equations are of the form \cite%
{Tamagaki,Takatsuka,Khod,Baldo,Elg,Khodel,Schwenk}: 
\begin{equation}
\Delta _{l,m_{j}}\left( p\right) =-\sum_{l^{\prime }}\frac{1}{2\varrho }\int
dp^{\prime }p^{\prime 2}i^{l-l^{\prime }}V_{ll^{\prime }}\left( p,p^{\prime
}\right) \Delta \left( p^{\prime }\right) \left\langle T\sum_{\eta }\frac{%
\mathbf{b}_{l^{\prime },m_{j}}^{\ast }(\mathbf{n}^{\prime })\mathbf{\bar{b}}(%
\mathbf{n}^{\prime })}{p_{\eta }^{2}+E_{\mathbf{p}^{\prime }}^{2}}%
\right\rangle .  \label{gap}
\end{equation}%
We are interested in the processes occurring in a vicinity of the Fermi
surface. To get rid of the integration over the regions far from the Fermi
surface we renormalize the interaction as suggested in Refs. \cite%
{Leggett,Leggett1}: we define 
\begin{equation}
V_{ll^{\prime }}^{\left( r\right) }\left( p,p^{\prime };T\right)
=V_{ll^{\prime }}\left( p,p^{\prime }\right) -\sum_{l^{\prime \prime }}\int 
\frac{dp^{\prime \prime }p^{\prime \prime 2}}{\pi ^{2}}V_{ll^{\prime \prime
}}\left( p,p^{\prime \prime }\right) \left( GG^{-}\right) _{N}^{\prime
\prime }V_{l^{\prime \prime }l^{\prime }}^{\left( r\right) }\left( p^{\prime
\prime },p^{\prime };T\right) \ ,  \label{Vr}
\end{equation}%
where the loop $\left( GG^{-}\right) _{n}$ is evaluated in the normal
(non-superfluid) state. In terms of $V_{ll^{\prime }}^{\left( r\right) }$
the renormalized gap equations can be written in the following matrix form 
\begin{equation}
\left( 
\begin{array}{c}
\Delta _{1,m_{j}} \\ 
\Delta _{3,m_{j}}%
\end{array}%
\right) =-\Delta \left( 
\begin{array}{cc}
V_{11}^{\left( r\right) } & -V_{13}^{\left( r\right) } \\ 
-V_{13}^{\left( r\right) } & V_{33}^{\left( r\right) }%
\end{array}%
\right) \left( 
\begin{array}{c}
\left\langle \mathbf{\bar{b}b}_{1,m_{j}}^{\ast }A\right\rangle  \\ 
\left\langle \mathbf{\bar{b}b}_{3,m_{j}}^{\ast }A\right\rangle 
\end{array}%
\right) ,  \label{gapeq}
\end{equation}%
assuming that in the narrow vicinity of the Fermi surface the smooth
functions $V_{ll^{\prime }}^{\left( r\right) }\left( p,p^{\prime }\right) $
and $\Delta \left( p^{\prime }\right) $ may be replaced with constants. In
obtaining Eq. (\ref{gapeq}) the fact is used that the interaction matrix is
symmetric on the Fermi surface, $V_{31}=V_{13}$. The function $A\left( 
\mathbf{n}\right) $ arises due to the renormalization procedure. It is given
by 
\begin{equation}
A\left( \mathbf{n}\right) =\frac{1}{2}\int_{0}^{\infty }d\varepsilon \left( 
\frac{1}{\sqrt{\varepsilon ^{2}+\Delta ^{2}\bar{b}^{2}}}\tanh \frac{\sqrt{%
\varepsilon ^{2}+\Delta ^{2}\bar{b}^{2}}}{2T}-\frac{1}{\varepsilon }\tanh 
\frac{\varepsilon }{2T}\right) .  \label{An}
\end{equation}

The interaction matrix can be diagonalized by unitary transformations $%
V^{^{\prime }}=UVU^{\dagger }$with $U$ being an unitary matrix 
\begin{equation}
U=\left( U^{-1}\right) ^{\dagger }=\frac{1}{\left( V_{+}+V_{-}\right) ^{%
\frac{1}{2}}\allowbreak }\left( 
\begin{array}{cc}
\sqrt{V_{+}} & \sqrt{V_{-}} \\ 
-\sqrt{V_{-}} & \sqrt{V_{+}}%
\end{array}%
\right) ,  \label{Umatr}
\end{equation}%
where$\allowbreak $ $V_{\pm }=\sqrt{\left( V_{33}^{\left( r\right)
}-V_{11}^{\left( r\right) }\right) ^{2}+4V_{13}^{\left( r\right) 2}}\pm
\left( V_{33}^{\left( r\right) }-V_{11}^{\left( r\right) }\right) $.

One has $UVU^{\dagger }=\mathrm{diag}\left( W_{-},W_{+}\right) $ with 
\begin{equation}
W_{\pm }=\frac{1}{2}\left( V_{11}^{\left( r\right) }+V_{33}^{\left( r\right)
}\pm \sqrt{\left( V_{33}^{\left( r\right) }-V_{11}^{\left( r\right) }\right)
^{2}+4V_{13}^{\left( r\right) 2}}\right) .  \label{W}
\end{equation}%
Applying the unitary transformation $U$\ to the gap equations (\ref{gapeq})
yields two coupled equations:%
\begin{eqnarray}
&&\sqrt{V_{+}}\Delta _{1,m_{j}}+\sqrt{V_{-}}\Delta _{3,m_{j}}  \notag \\
&=&-W_{-}\sum_{l}\Delta _{l,m_{j}}\left\langle \left( \sqrt{V_{+}}\mathbf{b}%
_{1,m_{j}}^{\ast }\mathbf{b}_{l,m_{j}}+\sqrt{V_{-}}\mathbf{b}%
_{3,m_{j}}^{\ast }\mathbf{b}_{l,m_{j}}\right) A\right\rangle ,  \label{Ge1}
\end{eqnarray}%
\begin{eqnarray}
&&\sqrt{V_{-}}\Delta _{1,m_{j}}-\sqrt{V_{+}}\Delta _{3,m_{j}}  \notag \\
&=&-W_{+}\sum_{l}\Delta _{l,m_{j}}\left\langle \left( \sqrt{V_{-}}\mathbf{b}%
_{1,m_{j}}^{\ast }\mathbf{b}_{l,m_{j}}-\sqrt{V_{+}}\mathbf{b}%
_{3,m_{j}}^{\ast }\mathbf{b}_{l,m_{j}}\right) A\right\rangle .  \label{Ge2}
\end{eqnarray}%
In obtaining these equations we made use of Eq. (\ref{bbar}) and
orthogonality relations (\ref{bbdfi}), assuming that the energy gap $\bar{b}%
\left( \mathbf{n}\right) \Delta $ is azimuth-symmetric \cite%
{Tamagaki,Takatsuka,Khod,Baldo,Elg,Khodel,Schwenk}.

We are interested in the linear medium response onto the external
axial-vector field. The field interaction with a superfluid should be
described with the aid of two ordinary and two anomalous three-point
effective vertices. In the BCS approximation, the ordinary axial-vector
vertices of a particle and a hole are to be taken as $\bm{\hat{\sigma}}$ and 
$\bm{\hat{\sigma}}^{T}$, respectively. The anomalous effective vertices, $%
\mathbf{\hat{T}}^{\left( 1\right) }\left( \mathbf{n;}\omega ,\mathbf{q}%
\right) $ and $\mathbf{\hat{T}}^{\left( 2\right) }\left( \mathbf{n;}\omega ,%
\mathbf{q}\right) $ are given by the infinite sums of the diagrams taking
account of the pairing interaction in the ladder approximation \cite{Larkin}%
. These $2\times 2$ vector matrices are to satisfy the Dyson's equations
symbolically depicted by graphs in Fig. \ref{fig1}. Analytic form of the
above diagrams is derived in Refs. \cite{L09a}. After some algebraic
manipulations the BCS equations for anomalous vertices can be found in the
following form (for brevity we omit the dependence of functions on $\omega $
and $\mathbf{q}$\textbf{)}:

\begin{figure}[h]
\includegraphics{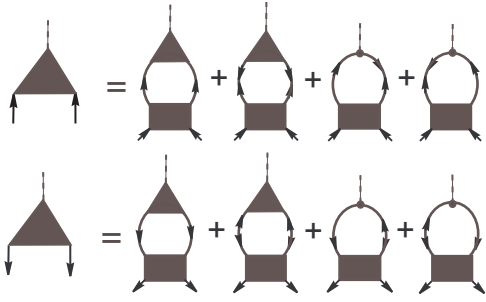}
\caption{Dyson's equations for the anomalous vertices. The shaded rectangle
represents the pairing interaction.}
\label{fig1}
\end{figure}

\begin{gather}
\mathbf{\hat{T}}^{\left( 1\right) }\left( \mathbf{n}\right) =\sum_{lm_{j}}%
\bm{\hat{\sigma}}\mathbf{b}_{lm_{j}}(\mathbf{n})\hat{g}\sum_{l^{\prime
}}V_{ll^{\prime }}\frac{1}{2}\left\langle \mathcal{I}_{GG^{-}}\mathrm{Tr}%
\left[ \hat{g}\left( \bm{\hat{\sigma}}\mathbf{b}_{l^{\prime }m_{j}}^{\ast
}\right) \mathbf{\hat{T}}^{\left( 1\right) }\right] \right.  \notag \\
\left. -\mathcal{I}_{FF}\mathrm{Tr}\left[ \left( \bm{\hat{\sigma}}\mathbf{b}%
_{l^{\prime }m_{j}}^{\ast }\right) \left( \bm{\hat{\sigma}}\mathbf{\bar{b}}%
\right) \hat{g}\mathbf{\hat{T}}^{\left( 2\right) }\left( \bm{\hat{\sigma}}%
\mathbf{\bar{b}}\right) \right] -\frac{\omega }{\Delta }\mathcal{I}%
_{FF}2i\left( \mathbf{b}_{l^{\prime }m_{j}}^{\ast }\mathbf{\times \bar{b}}%
\right) \right\rangle ,  \label{EqT1}
\end{gather}%
\begin{gather}
\mathbf{\hat{T}}^{\left( 2\right) }\left( \mathbf{n}\right) =\sum_{lm_{j}}%
\hat{g}\bm{\hat{\sigma}}\mathbf{b}_{lm_{j}}^{\ast }(\mathbf{n}%
)\sum_{l^{\prime }}V_{ll^{\prime }}\frac{1}{2}\left\langle \mathcal{I}%
_{G^{-}G}\mathrm{Tr}\left[ \left( \bm{\hat{\sigma}}\mathbf{b}_{l^{\prime
}m_{j}}\right) \hat{g}\mathbf{\hat{T}}^{\left( 2\right) }\right] \right. 
\notag \\
\left. -\mathcal{I}_{FF}\mathrm{Tr}\left[ \left( \bm{\hat{\sigma}}\mathbf{b}%
_{l^{\prime }m_{j}}\right) \left( \bm{\hat{\sigma}}\mathbf{\bar{b}}\right) 
\mathbf{\hat{T}}^{\left( 1\right) }\hat{g}\left( \bm{\hat{\sigma}}\mathbf{%
\bar{b}}\right) \right] -\frac{\omega }{\Delta }\mathcal{I}_{FF}2i\left( 
\mathbf{b}_{l^{\prime }m_{j}}\mathbf{\times \bar{b}}\right) \right\rangle .
\label{EqT2}
\end{gather}

Inspection of the equations reveals that the anomalous axial-vector vertices
can be found in the following form 
\begin{equation}
\mathbf{\hat{T}}^{\left( 1\right) }\left( \mathbf{n},\omega \right)
=\sum_{lm_{j}}\mathbf{B}_{l,m_{j}}\left( \omega \right) \left( %
\bm{\hat{\sigma}}\mathbf{b}_{l,m_{j}}\right) \hat{g},  \label{T1A}
\end{equation}%
\begin{equation}
\mathbf{\hat{T}}^{\left( 2\right) }\left( \mathbf{n},\omega \right)
=\sum_{lm_{j}}\mathbf{B}_{l,m_{j}}\left( \omega \right) \hat{g}\left( %
\bm{\hat{\sigma}}\mathbf{b}_{l,m_{j}}^{\ast }\right) .  \label{T2A}
\end{equation}%
As explained above we are interested in solutions with $\mathbf{q}=0$. Then
inserting of these forms into Eqs. (\ref{EqT1}), (\ref{EqT2}) allows to
obtain the equations for $\mathbf{B}_{l,m_{j}}\left( \omega \right) $. We
write the result in the matrix form (For brevity we omit the dependence on $%
\mathbf{n}$ and $\omega $)%
\begin{gather}
\left( 
\begin{array}{c}
\mathbf{B}_{1,m_{j}} \\ 
\mathbf{B}_{3,m_{j}}%
\end{array}%
\right) =-\left( 
\begin{array}{cc}
V_{11} & -V_{13} \\ 
-V_{13} & V_{33}%
\end{array}%
\right) \left\{ \left( 
\begin{array}{c}
\sum_{l}\left\langle \left( A+\frac{\omega ^{2}}{2\Delta ^{2}}\mathcal{I}%
_{0}\right) \left( \mathbf{b}_{1,m_{j}}^{\ast }\mathbf{b}_{l,m_{j}}\right)
\right\rangle \mathbf{B}_{l,m_{j}} \\ 
\sum_{l}\left\langle \left( A+\frac{\omega ^{2}}{2\Delta ^{2}}\mathcal{I}%
_{0}\right) \left( \mathbf{b}_{3,m_{j}}^{\ast }\mathbf{b}_{l,m_{j}}\right)
\right\rangle \mathbf{B}_{l,m_{j}}%
\end{array}%
\right) \right.  \notag \\
\left. -2\left( 
\begin{array}{c}
\sum_{l}\left\langle \mathcal{I}_{0}\left( \mathbf{\bar{b}b}_{1,m_{j}}^{\ast
}\right) \left( \mathbf{\bar{b}b}_{l,m_{j}}\right) \right\rangle \mathbf{B}%
_{l,m_{j}} \\ 
\sum_{l}\left\langle \mathcal{I}_{0}\left( \mathbf{\bar{b}b}_{3,m_{j}}^{\ast
}\right) \left( \mathbf{\bar{b}b}_{l,m_{j}}\right) \right\rangle \mathbf{B}%
_{l,m_{j}}%
\end{array}%
\right) +\frac{\omega }{\Delta }i\left( 
\begin{array}{c}
\left\langle \mathcal{I}_{0}\left( \mathbf{b}_{1,m_{j}}^{\ast }\mathbf{%
\times \bar{b}}\right) \right\rangle \\ 
\left\langle \mathcal{I}_{0}\left( \mathbf{b}_{3,m_{j}}^{\ast }\mathbf{%
\times \bar{b}}\right) \right\rangle%
\end{array}%
\right) \right\} .  \label{B}
\end{gather}

In this equation, the interaction matrix can be diagonalized by the unitary
transformation (\ref{Umatr}). Further simplification is possible due to the
fact that by virtue of Eqs. (\ref{Ge1}), (\ref{Ge2}) the coupling constants $%
W_{\pm }$ can be removed out of the equations. Explicit evaluation of
equations obtained in this way for arbitrary values of $\omega $ and $T$
requires numerical computation. However, we can get a clear idea of the
behavior of the vertex functions using the angle-averaged energy gap $\Delta
^{2}\bar{b}^{2}\rightarrow \left\langle \Delta ^{2}\bar{b}^{2}\right\rangle
=\Delta ^{2}$ in the quasiparticle energy $E_{\mathbf{p}}$. In this
approximation, the functions $\mathcal{I}\left( \omega ,T\right) $ and $%
A\left( T\right) $ can be moved beyond the angle integrals. Performing
trivial integrations we then get a set of linear equations (two equations
for each value of $m_{j})$. It is convenient to denote%
\begin{equation}
\beta _{l,l^{\prime }}^{\left( m_{j}\right) }\equiv \left\langle \left( 
\mathbf{b}_{l,m_{j}}\mathbf{\bar{b}}\right) \left( \mathbf{b}_{l^{\prime
},m_{j}}^{\ast }\mathbf{\bar{b}}\right) \right\rangle ,  \label{beta}
\end{equation}%
and 
\begin{equation}
\Omega =\frac{\omega }{2\Delta }.  \label{OMEGA}
\end{equation}%
Then the set of equations can be written in the form$~$%
\begin{gather}
\mathbf{B}_{1,m_{j}}\left[ \sqrt{V_{+}}\left( \Omega ^{2}-\beta
_{1,1}^{\left( m_{j}\right) }\right) -\sqrt{V_{-}}\beta _{1,3}^{\left(
m_{j}\right) }\right]  \notag \\
+\mathbf{B}_{3,m_{j}}\left[ \sqrt{V_{-}}\left( \Omega ^{2}-\beta
_{3,3}^{\left( m_{j}\right) }\right) -\sqrt{V_{+}}\beta _{3,1}^{\left(
m_{j}\right) }\right]  \notag \\
=-i\Omega \left( \sqrt{V_{+}}\left\langle \mathbf{b}_{1,m_{j}}^{\ast }\times 
\mathbf{\bar{b}}\right\rangle +\sqrt{V_{-}}\left\langle \mathbf{b}%
_{3,m_{j}}^{\ast }\times \mathbf{\bar{b}}\right\rangle \right) ,
\label{eqB1}
\end{gather}%
\begin{gather}
\mathbf{B}_{1,m_{j}}\left[ -\sqrt{V_{-}}\left( \Omega ^{2}-\beta
_{1,1}^{\left( m_{j}\right) }\right) -\sqrt{V_{+}}\beta _{1,3}^{\left(
m_{j}\right) }\right]  \notag \\
+\mathbf{B}_{3,m_{j}}\left[ \sqrt{V_{+}}\left( \Omega ^{2}-\beta
_{3,3}^{\left( m_{j}\right) }\right) +\sqrt{V_{-}}\beta _{3,1}^{\left(
m_{j}\right) }\right]  \notag \\
=-i\Omega \left( -\sqrt{V_{-}}\left\langle \mathbf{b}_{1,m_{j}}^{\ast
}\times \mathbf{\bar{b}}\right\rangle +\sqrt{V_{+}}\left\langle \mathbf{b}%
_{3,m_{j}}^{\ast }\times \mathbf{\bar{b}}\right\rangle \right) ,
\label{eqB2}
\end{gather}%
which can be solved to give $\allowbreak $%
\begin{equation}
\mathbf{B}_{1,m_{j}}=\frac{-i\Omega }{\chi }\left[ \left( \Omega ^{2}-\beta
_{3,3}^{\left( m_{j}\right) }\right) \left\langle \mathbf{b}_{1,m_{j}}^{\ast
}\times \mathbf{\bar{b}}\right\rangle +\beta _{3,1}^{\left( m_{j}\right)
}\left\langle \mathbf{b}_{3,m_{j}}^{\ast }\times \mathbf{\bar{b}}%
\right\rangle \right] ,  \label{B1m}
\end{equation}%
\begin{equation}
\mathbf{B}_{3,m_{j}}=\frac{-i\Omega }{\chi }\left[ \left( \Omega ^{2}-\beta
_{1,1}^{\left( m_{j}\right) }\right) \left\langle \mathbf{b}_{3,m_{j}}^{\ast
}\times \mathbf{\bar{b}}\right\rangle +\beta _{1,3}^{\left( m_{j}\right)
}\left\langle \mathbf{b}_{1,m_{j}}^{\ast }\times \mathbf{\bar{b}}%
\right\rangle \right] ~  \label{B3m}
\end{equation}%
with 
\begin{equation}
\chi \left( \Omega \right) \equiv \Omega ^{4}-\Omega ^{2}\left( \beta
_{1,1}^{\left( m_{j}\right) }+\beta _{3,3}^{\left( m_{j}\right) }\right)
+\beta _{1,1}^{\left( m_{j}\right) }\beta _{3,3}^{\left( m_{j}\right)
}-\beta _{1,3}^{\left( m_{j}\right) }\beta _{3,1}^{\left( m_{j}\right) }.
\label{hiw}
\end{equation}

As is well known, poles of the vertex function correspond to collective
eigen-modes of the system. Eigen- frequencies, $\Omega =\Omega ^{\left(
m_{j}\right) }$, of such oscillations satisfy the equation $\chi \left(
\Omega ^{\left( m_{j}\right) }\right) =0$. This equation gives%
\begin{equation}
\left( \Omega _{\pm }^{\left( m_{j}\right) }\right) ^{2}=\frac{1}{2}\left(
\beta _{1,1}^{\left( m_{j}\right) }+\beta _{3,3}^{\left( m_{j}\right) }\pm 
\sqrt{\left( \beta _{1,1}^{\left( m_{j}\right) }-\beta _{3,3}^{\left(
m_{j}\right) }\right) ^{2}+4\beta _{1,3}^{\left( m_{j}\right) }\beta
_{3,1}^{\left( m_{j}\right) }}\right) .\allowbreak   \label{Omega}
\end{equation}%
\ Notice that the interaction parameters,$V_{\pm }$, drop out of the above
solutions, which depend explicitly only on the partial gap amplitudes. This
means that\ the contribution of excited bound pairs with $l=3$ into the spin
oscillations is caused basically by spin-orbit interactions but not by the
tensor forces.

Indeed, in Eqs. (\ref{B1m}) - (\ref{Omega}), the equilibrium order parameter
is specified solely by means of the real vector $\mathbf{\bar{b}}$. If we
switch off the interaction in the $^{3}F_{2}$ and $^{3}P_{2}-$\noindent $%
^{3}F_{2}$ channels and consider pure $^{3}P_{2}$ pairing with $m_{j}=0$ we
are then left with $\mathbf{\bar{b}=b}_{1,0}$ and $\Delta =\Delta _{1,0}$. \
In this case, in Eqs. (\ref{B1m}), (\ref{B3m}), one has: 
\begin{equation}
\int \frac{d\varphi }{2\pi }\left( \mathbf{b}_{1,m_{j}}^{\ast }\times 
\mathbf{\bar{b}}\right) =\int \frac{d\varphi }{2\pi }\left( \mathbf{b}%
_{3,m_{j}}^{\ast }\times \mathbf{\bar{b}}\right) =0~~\mathsf{for}%
~~m_{j}=0,\pm 2,  \label{cross}
\end{equation}%
and the non-trivial solutions exist only for $m_{j}=\pm 1$. The explicit
form of $\mathbf{b}_{l,m_{j}}$ can be obtained from Eq. (\ref{blm}): 
\begin{equation}
\mathbf{b}_{1,0}=\sqrt{\frac{1}{2}}\left( 
\begin{array}{c}
-n_{1} \\ 
-n_{2} \\ 
2n_{3}%
\end{array}%
\right) \,,~\mathbf{b}_{1,1}=-\sqrt{\frac{3}{4}}\left( 
\begin{array}{c}
n_{3} \\ 
in_{3} \\ 
n_{1}+in_{2}%
\end{array}%
\right) ,  \label{b1mj}
\end{equation}%
\begin{equation}
\mathbf{b}_{3,0}=\sqrt{\frac{3}{4}}\left( 
\begin{array}{c}
n_{1}\left( 1-5n_{3}^{2}\right)  \\ 
n_{2}\left( 1-5n_{3}^{2}\right)  \\ 
n_{3}\left( 3-5n_{3}^{2}\right) 
\end{array}%
\right) \,,~\mathbf{b}_{3,1}=\sqrt{\frac{1}{2}}\left( 
\begin{array}{c}
n_{3}\left( 1-5n_{1}\left( n_{1}+in_{2}\right) \right)  \\ 
in_{3}\left( 1+5in_{2}\left( n_{1}+in_{2}\right) \right)  \\ 
\left( n_{1}+in_{2}\right) \left( 1-5n_{3}^{2}\right) 
\end{array}%
\right) .  \label{b3mj}
\end{equation}%
Making use of these expressions in Eq. (\ref{beta}) we find 
\begin{equation}
\beta _{1,1}^{\left( \pm 1\right) }=\frac{1}{20}~,~\beta _{3,3}^{\left( \pm
1\right) }=\frac{29}{70}~,~\beta _{1,3}^{\left( \pm 1\right) }=\beta
_{3,1}^{\left( \pm 1\right) }=-\frac{1}{70}\sqrt{\frac{3}{2}}.
\label{betamj}
\end{equation}%
Inserting these values into Eq. (\ref{Omega}) we find $4\beta _{1,3}^{\left(
\pm 1\right) }\beta _{3,1}^{\left( \pm 1\right) }\ll \left( \beta
_{3,3}^{\left( \pm 1\right) }-\beta _{1,1}^{\left( \pm 1\right) }\right) ^{2}
$. By neglecting the small term $4\beta _{1,3}^{\left( \pm 1\right) }\beta
_{3,1}^{\left( \pm 1\right) }$ under the root in Eq. (\ref{Omega}) we obtain
two (twofold) eigen-frequencies of spin oscillations in the condensate with $%
m_{j}=\pm 1$: 
\begin{equation}
\omega _{-}^{\left( m_{j}\right) }~\simeq 2\Delta \sqrt{\beta _{1,1}^{\left(
\pm 1\right) }}=\frac{1}{\sqrt{5}}\Delta ,  \label{w1p}
\end{equation}%
\begin{equation}
\omega _{+}^{\left( m_{j}\right) }\simeq 2\Delta \sqrt{\beta _{3,3}^{\left(
\pm 1\right) }}=\sqrt{\allowbreak \frac{58}{35}}\Delta .  \label{w2p}
\end{equation}

In Refs. \cite{L10a,L10b}, eigen-modes of spin oscillations in the $^{3}P_{2}
$ superfluid neutron liquid was studied in a simple model restricted to
excitations of the condensate with $l=1$. The spin wave energy (at $\mathbf{q%
}=0$) was found to be $\omega _{m_{j}}=\Delta /\sqrt{5}$. Equations (\ref%
{Omega}), (\ref{w1p}), (\ref{w2p}) show that extending of the decomposition
up to $l=1,3$ in Eqs. (\ref{T1A}), (\ref{T2A}) leads to a very small
frequency shift of the known mode, $\omega =\omega _{-}^{\left( m_{j}\right)
}\simeq \omega _{m_{j}}$, but opens the new additional mode of spin
oscillations with $\omega =\omega _{+}^{\left( m_{j}\right) }$.

\begin{figure}[h]
\includegraphics{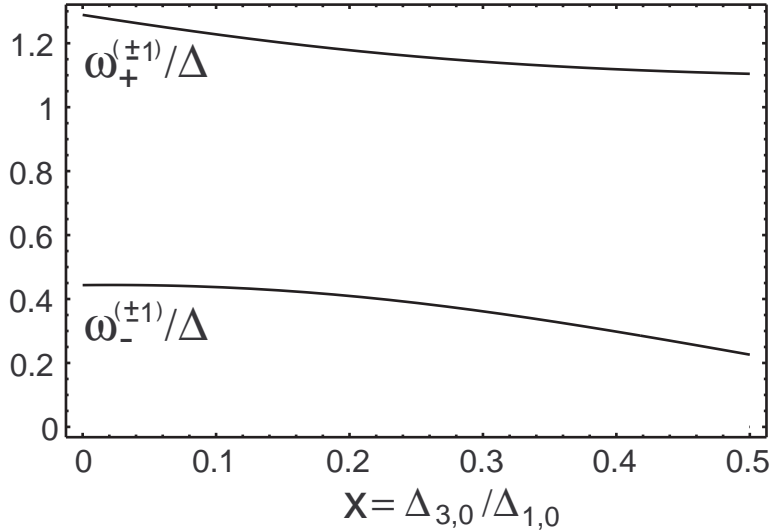}
\caption{The energy gaps for the collective spin excitations $\protect\omega %
_{-}^{\left( m_{j}\right) }$ and $\protect\omega _{+}^{\left( m_{j}\right) }$
vs the ratio of partial gap amplitudes in the $^{3}F_{2}$ and $^{3}P_{2}$
channels. The energy gap of a neutron quasiparticle is given by $\Delta
^{2}=\Delta _{1,0}^{2}+\Delta _{3,0}^{2}$.}
\label{fig2}
\end{figure}

Neutrino decays of spin waves can play an important role in the cooling
scenario of neutron stars. A simple estimate made in Ref. \cite{L10b} has
shown that the decays of spin waves with $\omega _{m_{j}}=\Delta /\sqrt{5}~$%
can become the dominant cooling mechanism in a wide range of low
temperatures and modify the cooling trajectory of neutron stars. As well as
the first mode, the second mode of spin oscillations is kinematically able
to decay into neutrino pair. Therefore let us examine the wave excitation
energies more accurately with taking into account the tensor forces. We will
again focus on the condensation with $m_{j}=0~$by assuming $\Delta
^{2}=\Delta _{1,0}^{2}+\Delta _{3,0}^{2}$, and 
\begin{equation}
\mathbf{\bar{b}}\left( \mathbf{n}\right) =\frac{\Delta _{1,0}}{\Delta }%
\mathbf{b}_{1,0}\left( \mathbf{n}\right) +\frac{\Delta _{3,0}}{\Delta }%
\mathbf{b}_{3,0}\left( \mathbf{n}\right) .  \label{bbarPF}
\end{equation}%
In this case Eqs. (\ref{cross}) are still valid and the non-trivial
solutions to Eqs. (\ref{B1m}), (\ref{B3m}) exist only for $m_{j}=\pm 1$.
Insertion of the expression (\ref{bbarPF}) into Eq. (\ref{beta}) results in 
\begin{eqnarray}
\beta _{1,1}^{\left( \pm 1\right) } &=&\frac{1}{20}\frac{\Delta _{1}^{2}}{%
\Delta ^{2}}\left( 1-\frac{2}{7}\sqrt{6}\frac{\Delta _{3}}{\Delta _{1}}+%
\frac{58\Delta _{3}^{2}}{7\Delta _{1}^{2}}\right) ,  \notag \\
\beta _{3,3}^{\left( \pm 1\right) } &=&\frac{29}{70}\frac{\Delta _{1}^{2}}{%
\Delta ^{2}}\left( 1-\frac{32}{87}\sqrt{6}\frac{\Delta _{3}}{\Delta _{1}}+%
\frac{28}{87}\frac{\Delta _{3}^{2}}{\Delta _{1}^{2}}\right) ,  \notag \\
\beta _{1,3}^{\left( \pm 1\right) } &=&\beta _{3,1}^{\left( \pm 1\right) }=-%
\frac{1}{140}\sqrt{6}\frac{\Delta _{1}^{2}}{\Delta ^{2}}\left( 1-11\sqrt{6}%
\frac{\Delta _{3}}{\Delta _{1}}+\frac{32}{3}\frac{\Delta _{3}^{2}}{\Delta
_{1}^{2}}\right) .  \label{beta31}
\end{eqnarray}%
Because $\beta _{l,l^{\prime }}^{\left( 1\right) }=\beta _{l,l^{\prime
}}^{\left( -1\right) }\equiv \beta _{l,l^{\prime }}$ we further omit the
superscript $m_{j}=\pm 1$ by assuming that all the frequencies are twofold.
Making use of Eqs. (\ref{beta31}) we find 
\begin{align}
\omega _{-}^{2}& =\Delta _{1,0}^{2}\left( \frac{13}{14}-\frac{1}{3}\sqrt{6}%
\frac{\Delta _{3}}{\Delta _{1}}+\frac{23}{21}\frac{\Delta _{3}^{2}}{\Delta
_{1}^{2}}\right.   \notag \\
& \left. -\sqrt{\frac{15}{28}-\frac{25}{49}\sqrt{6}\frac{\Delta _{3}}{\Delta
_{1}}+\frac{485}{147}\frac{\Delta _{3}^{2}}{\Delta _{1}^{2}}-\allowbreak 
\frac{370}{441}\sqrt{6}\frac{\Delta _{3}^{3}}{\Delta _{1}^{3}}+\frac{55}{63}%
\frac{\Delta _{3}^{4}}{\Delta _{1}^{4}}}\right) ,  \label{w3}
\end{align}%
\begin{align}
\omega _{+}^{2}& =\Delta _{1,0}^{2}\left( \frac{13}{14}-\frac{1}{3}\sqrt{6}%
\frac{\Delta _{3}}{\Delta _{1}}+\frac{23}{21}\frac{\Delta _{3}^{2}}{\Delta
_{1}^{2}}\right.   \notag \\
& \left. +\sqrt{\frac{15}{28}-\frac{25}{49}\sqrt{6}\frac{\Delta _{3}}{\Delta
_{1}}+\frac{485}{147}\frac{\Delta _{3}^{2}}{\Delta _{1}^{2}}-\allowbreak 
\frac{370}{441}\sqrt{6}\frac{\Delta _{3}^{3}}{\Delta _{1}^{3}}+\frac{55}{63}%
\frac{\Delta _{3}^{4}}{\Delta _{1}^{4}}}\right) .  \label{w4}
\end{align}%
In Fig. \ref{fig2}, the energy of the collective spin excitations (at $%
\mathbf{q}=0$) is shown vs the ratio of the partial gaps $x=\Delta
_{3,0}/\Delta _{1,0}$. 
According calculations of different authors, at the Fermi surface one has $%
\Delta _{3}\simeq 0.17\Delta _{1}$ (see, \textit{e.g.}, Ref. \cite{Khod}).
In this case our theoretical analysis predicts two degenerate modes with $%
\omega =\omega _{-}=0.42\Delta $, and two degenerate modes with $\omega
=\omega _{+}=1.\,\allowbreak 19\Delta $.

Because of a rather small excitation energy the decay of the corresponding
collective spin excitations into neutrino pairs should lead to an extension
of the low-temperature domain where the volume neutrino emission dominates
the surface gamma radiation in the star cooling. This effect was already
demonstrated in Ref. \cite{L10b}, where only the lowest branch of the
collective spin excitations $\omega =\Delta /\sqrt{5}\simeq \omega _{-}$ has
been taken into account. The neutrino emissivity caused by the decay of the
new spin modes predicted in this paper will be considered elsewhere.

\end{document}